\def \be {\begin{equation}}

\def \half{\frac{1}{2}}

\def \real {{\bf R}}
\def \rf {(\ref}
\def \eqn {equation}
\def \sln {solution}

\def \tfn {transformation}
\def \mi {matri}
\def\be{\begin{equation}}
\def\ee{\end{equation}}
\def\ll{\label}

\documentstyle[12pt]{article}
{\LARGE \bf\author{{\sc  Ladislav   Hlavat\'{y}}
\thanks{Postal   address:
B\v{r}ehov\'{a} 7, 115 19
Prague 1, Czech Republic. E-mail: hlavaty@br.fjfi.cvut.cz}
\\ {\it Department  of  Physics,}
\\ {\it  Faculty  of  Nuclear  Sciences  and
Physical Engineering}}
\title{All generalized SU(2) chiral models have spectral dependent Lax
formulation}
\begin{document}
\maketitle
\abstract{The equations that define the Lax pairs for generalized principal
chiral models can be solved for any nondegenerate bilinear form on $su(2)$.
The \sln{} is dependent on one free variable that can serve as the spectral
parameter.}

\section {Formulation of the generalized chiral models}
Principal chiral models are important example of relativistically invariant
field theory integrable by the inverse scattering method. Their Lax pair
formulation was given in \cite{zami:relinm}.
Generalized principal chiral models were introduced as a quasiclassical
limit of the Baxter quantum XYZ model in \cite{cher:relinm}. The equations
of motion were written in the zero curvature form but without a spectral
parameter. The purpose of this paper is to introduce the spectral dependent
Lax pair for general $SU(2)$ model.

Generalized principal chiral models on the Lie group $G$ are defined by the
action
\be I[g]=-\int d^2x \eta^{\mu\nu}L(A_\mu,A_\nu),
\ll{invaction}\ee
where\be A_{\mu}:=-i(g^{-1}\partial_\mu g), \ll{amu}\ee
$g:\real^2\rightarrow G,\ \mu,\nu\in\{0,1\},\ \eta:=diag(1,-1)$
and $L$ is a bilinear form on the corresponding Lie algebra. In the
coordinate dependent version
\be I[g]=-\int d^2x L_{ab}\eta^{\mu\nu}(g^{-1}\partial_\mu g)^a
(g^{-1}\partial_\nu g)^b,
\ll{action}\ee
where $L_{ab}$ is matrix $dim G\times dim G$ defined by the bilinear form
\be L_{ab}:=L(t_a,t_b), \
\ee
and $t_b$ are elements of basis in the Lie algebra
\be iA_\nu=(g^{-1}\partial_\nu g)=(g^{-1}\partial_\nu g)^bt_b= iA_\nu^bt_b.
\ee
Lie products of elements of basis define the structure coefficients
\be [t_a,t_b]=i{f_{ab}}^ct_c
\ll{crfort}\ee

The \eqn s of motion for the models read
\be \partial_\mu A^{\mu,a}+{S^a}_{bc}A_\mu^bA^{\mu,c}=0
\ll{eqnsmot}\ee
where
\be {S^a}_{bc}:=\half({F^a}_{bc}+{F^a}_{cb}),\ {F^a}_{bc}:=L_{qb}
{f_{cp}}^q(L^{-1})^{pa}.
\ll{sdef}\ee

\section{The Lax pairs}
In the paper \cite{soch:igpcm},  ansatz for the Lax formulation of the
generalized chiral models was taken in the form
\be [\partial_0+ N_{0,b}^a A_0^b t_a+ N_{1,b}^a A_1^b t_a\ ,\
     \partial_1+ N_{0,b}^a A_1^b t_a+ N_{1,b}^a A_0^b t_a]=0
\ll{lpansatz}\ee
where $N_{\mu}$ are two auxiliary $dim G\times dim G$ \mi ces.
It is known that the ansatz \rf{lpansatz}) provides Lax pairs for $L$
proportional to the Killing form $\kappa_{ab}:=Tr(t_at_b)$ and for the
anisotropic $SU(2)$ model where $L_{ab}:=L_a\kappa_{ab}$ (no summation) with
$L_1=L_2\neq L_3$.
We are going to prove that this ansatz gives Lax pairs for any bilinear
nondegenerate form $L$ on $su(2)$.

Necessary conditions that 
the operators in \rf{lpansatz}) form the Lax pair for the \eqn s of motion
\rf{eqnsmot}) are
\be N_{0,b}^a {f_{pq}}^b + iN_{\mu,p}^b{N^{\mu,c}}_{q} {f_{bc}}^a=0,
\ll{condN}\ee
\be \frac{i}{2}{f_{cd}}^a(N_{0,p}^c N_{1,q}^d+N_{0,q}^c N_{1,p}^d)
= N_{1,b}^a {S^b}_{pq}.
\ll{condNS}\ee
If $N_1$ is invertible, that we shall assume in the following, then these
conditions are also sufficient.
Note that the first condition is independent of the bilinear form $L$ so
that one can start with solving the \eqn{} \rf{condN}) and then look for the
bilinear form $L$ determined by \rf{condNS}) and \rf{sdef}).

The structure coefficients for $su(2)$ can be chosen in terms of the totally
antisymetric Levi-Civita tensor
\be {f_{ab}}^c=\epsilon_{abc}. \ll{feps}\ee
On the other hand, as the \mi x $L_{ab}$ is symmetric we can diagonalize it
by orthogonal transformations.
The important fact is that the structure coefficients \rf{feps}) remain
invariant under these transformations. It means that without loss of
generality for the $SU(2)$ models we can assume that $L$ is diagonal and
structure coefficients  have the form \rf{feps}).
We shall prove that we can satisfy the conditions \rf{condN}), \rf{condNS})
for any diagonal $L$ by the diagonal \mi ces $N_0$ and $N_1$ containing one
free ("spectral") parameter.

Inserting \rf{feps}) and
\be L=diag(L_1,L_2,L_3)
\ll{diagL}\ee
into \rf{sdef}) we find that the only nonzero components of $S^a_{bc}$ are
\[ S^1_{23}=S^1_{32}=\frac{L_2-L_3}{2L_1}=:\sigma_1/2,\]
\be S^2_{13}=S^2_{31}=\frac{L_3-L_1}{2L_2}=:\sigma_2/2,
\ll{sigdef}\ee
\[ S^3_{21}=S^3_{12}=\frac{L_1-L_2}{2L_3}=:\sigma_3/2.\]
The conditions \rf{condN}), \rf{condNS}) for
\be N_0=diag(P_1,P_2,P_3),\ N_1=diag(Q_1,Q_2,Q_3)
\ll{diagN}\ee
provide us with two sets  of equations for $P_j$ and $Q_j$, namely
\be Q_1Q_2=P_1P_2-iP_3,\ {\rm and\ cyclic\ permutations\ of}\ (1,2,3)
\ll{qpeqn}\ee
and
\be \sigma_1Q_1+iP_3Q_2-iP_2Q_3=0,\ {\rm and\ cyclic\ permutations\ of}\
(1,2,3).
\ll{qpsigeqn}\ee

Assuming that $N_1$ is invertible, i.e. $Q_j\neq 0$, we get  solution of the
\eqn s \rf{qpeqn})
\be Q_1=\pm \sqrt{\frac{R_2R_3}{R_1} },
\ Q_2=\frac{R_3}{Q_1},
\ Q_3=\frac{R_1Q_1}{Q_3}.
\ll{qsoln}\ee
where
\be R_1:=P_2P_3-iP_1,\ R_2:=P_3P_1-iP_2,\ R_3:=P_1P_2-iP_3.
\ll{rdef}\ee
The \sln{} is unique up to the sign.

Inserting (\ref{qsoln}),\rf{rdef}) into \rf{qpsigeqn}) we get three
nonlinear nonhomogeneous \eqn s for $P_j$
\be
E_1:=P_1P_2[\sigma_3(P_3^2-1)+P_2^2-P_1^2] -iP_3[(P_1^2+P_2^2)\sigma_3+P_2^2
-P_1^2]=0,
\ll{peqn3}\ee
\be
E_2:=P_2P_3[\sigma_1(P_1^2-1)+P_3^2-P_2^2] -iP_1[(P_2^2+P_3^2)\sigma_1+P_3^2
-P_2^2]=0,
\ll{peqn1}\ee
\be
_3:=P_3P_1[\sigma_2(P_2^2-1)+P_1^2-P_3^2]  -iP_2[(P_3^2+P_1^2)\sigma_2+P_1^2
-P_3^2]=0.
\ll{peqn2}\ee
These equations are not independent because the following relation holds
identically
\[ E_1P_1L_1(-L_1+L_2+L_3)+ E_2P_2L_2(-L_2+L_3+L_1)\]
\be + E_3P_3L_3(-L_3+L_1+L_2)=0.
\ee
That's why the variety of solutions of \rf{peqn3})--\rf{peqn2})  has the
dimension one. The \sln{} curve  can be written as
\[ P_1=i\kappa_1\frac{\sqrt{\mu +L_2}\sqrt{\mu +L_3}}
{\sqrt{L_2}\sqrt{L_3}},\
P_2=i\kappa_2\frac{\sqrt{\mu +L_3}\sqrt{\mu +L_1}}
{\sqrt{L_3}\sqrt{L_1}},\]
\be P_3=i\kappa_3\frac{\sqrt{\mu +L_1}\sqrt{\mu +L_2}}
{\sqrt{L_1}\sqrt{L_2}}
\ll{psoln} \ee
where $\mu$ is a free  parameter and
\[ \kappa_1^2=\kappa_2^2=1,\  \kappa_3= \kappa_1\kappa_2. \]
Inserting \rf{psoln}) into \rf{qsoln}) we get
\be Q_1=i\omega_1\frac{\sqrt{\mu}\sqrt{\mu +L_1}}
{\sqrt{L_2}\sqrt{L_3}},\
Q_2=i\omega_2\frac{\sqrt{\mu}\sqrt{\mu +L_2}}
{\sqrt{L_3}\sqrt{L_1}},\
Q_3=i\omega_3\frac{\sqrt{\mu}\sqrt{\mu +L_3}}
{\sqrt{L_1}\sqrt{L_2}},
\ll{qsolnlam} \ee
where
\[ \omega_1=\kappa_2\omega_3,\ \omega_2=\kappa_1\omega_3,
\ \omega_3^2=1. \]

The formulas \rf{diagN}), \rf{psoln}), and \rf{qsolnlam}) yield the \sln{}
of the \eqn s \rf{condN}), \rf{condNS}) for
\be L_{ab}=L_a\delta_{ab},\ {f_{ab}}^c=\epsilon_{abc}
\ee
and up to an eventual orthogonal \tfn{} of the algebra basis, they define
the Lax pair for the general principal $SU(2)$ chiral model \rf{action}).
It is easy to check that  for $L_1=L_2=L_3$ and $L_1=L_2\neq L_3$, the Lax
pairs coincide with the previously known cases \cite{zami:relinm},
\cite{soch:igpcm}.

The author gladly acknowledge assistence of the system Reduce \cite{red36}
by the presented calculations.

\end{document}